\documentclass[prb,twocolumn,showpacs,preprintnumbers,amsmath,amssymb]{revtex4}
\usepackage{graphicx}

\begin{document}
\preprint{}

\title{Structural and dielectric properties of Sr$_{2}$TiO$_{4}$ from first principles}
\author{Craig J. Fennie and Karin M. Rabe}
\affiliation{Department of Physics and Astronomy, Rutgers University,
        Piscataway, NJ 08854-8019}
\date{\today}

\begin{abstract}
We have investigated the structural and dielectric properties of Sr$_{2}$TiO$_{4}$,
the first member of the Sr$_{n+1}$Ti$_{n}$O$_{3n+1}$ Ruddlesden-Popper
series, within density functional theory. Motivated by recent work in which thin 
films of Sr$_{2}$TiO$_{4}$ were grown by molecular beam epitaxy (MBE) on SrTiO$_{3}$ 
substrates, the in-plane lattice parameter was fixed to the theoretically optimized
lattice constant of cubic SrTiO$_{3}$ (n=$\infty$), while the out-of-plane lattice 
parameter and the internal structural parameters were relaxed. The fully relaxed structure was 
also investigated. Density functional perturbation theory was used to calculate
the zone-center phonon frequencies, Born effective charges, and the electronic 
dielectric permittivity tensor. A detailed study of the contribution of individual 
infrared-active modes to the static dielectric permittivity tensor was performed. 
The calculated Raman and infrared phonon frequencies were found to be in agreement
with experiment where available. Comparisons of the calculated static dielectric 
permittivity with experiments on both ceramic powders and epitaxial thin films are 
discussed.
\end{abstract}


\maketitle

\section{Introduction}
\label{sec:Intro}
\begin{figure*}
\includegraphics[scale=0.6]{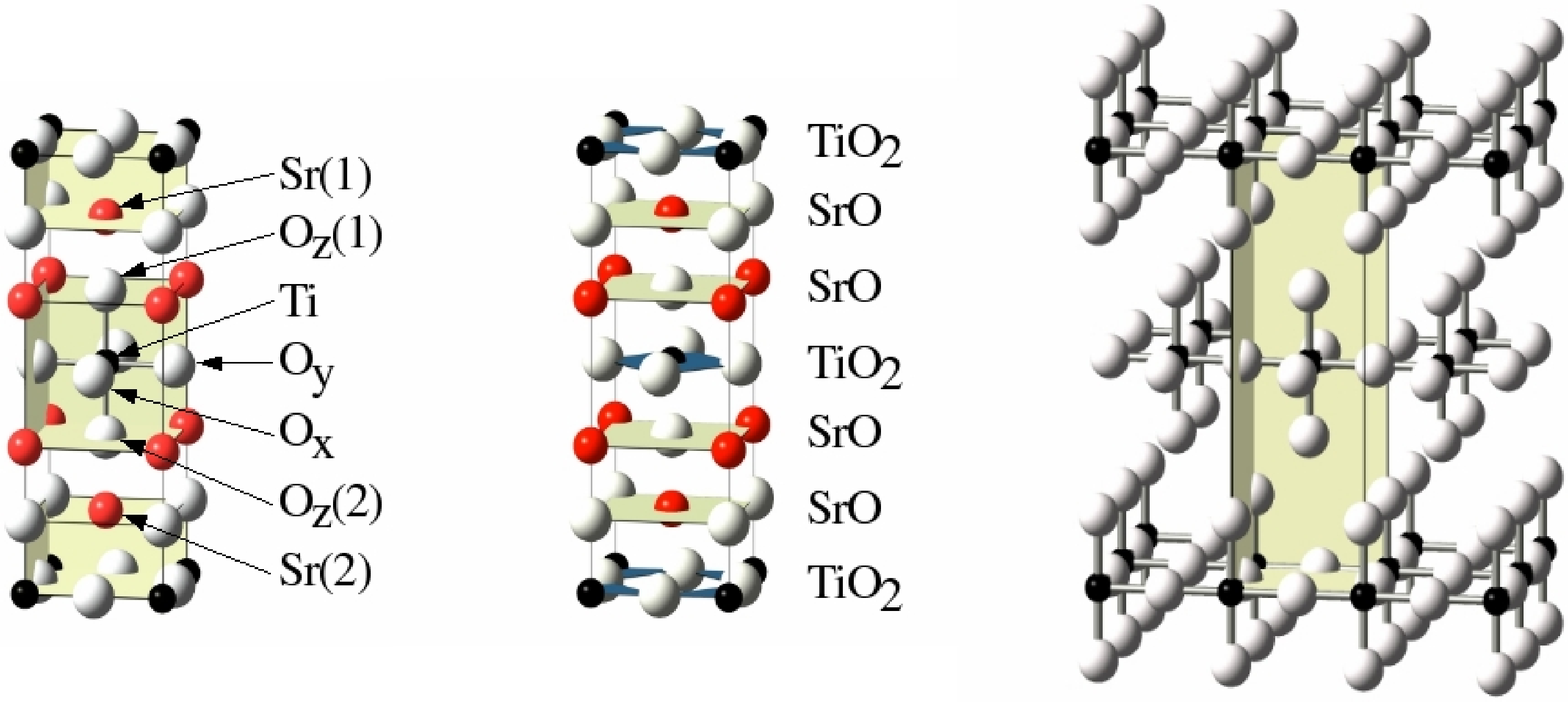}\\[0.3cm]
\caption{\label{fig:struct}
(Color online) The structure of Sr$_{2}$TiO$_{4}$ (space group $I4/mmm$) can be viewed as (a) 
a stacking of SrO-terminated SrTiO$_{3}$ perovskite [001] slabs, (b) a stacking 
of TiO$_{2}$ and of SrO planes along [001], and (c) a series of Ti-O chains, 
infinitely long in the plane along [100] and [010], and of finite extent along 
[001] (the Sr-atoms have been removed for clarity).}
\end{figure*}

Nonstoichiometric SrTiO$_{3}$ resists the formation of point defects by 
forming crystallographic shear (CS) phases.~\cite{Tilley} For excess SrO, 
these CS phases form the Sr$_{n+1}$Ti$_{n}$O$_{3n+1}$ Ruddlesden-Popper (RP) 
series. The structures of the members of the RP series can be viewed as a 
stacking of SrO-terminated SrTiO$_{3}$ perovskite [001] slabs with relative 
shifts of ($a_0$/2)[110], the slabs in the $n$th member of the series having 
a thickness of $n$ cubic perovskite lattice constants. This representation of 
the structure for $n$ = 1 is shown in Fig.~\ref{fig:struct}(a).

Considerable fundamental and practical interest in the RP series arises from the
dielectric properties of its end member SrTiO$_{3}$ ($n=\infty$). Strontium titanate 
is an incipient ferrroelectric with a high dielectric constant that can be readily 
tuned by applying a small DC bias.  SrTiO$_{3}$ has low dielectric loss at microwave 
frequencies but a large temperature coefficient of dielectric constant (TCF). 
The prospect of ``engineering" the properties of the constituent SrTiO$_3$ layers by 
varying $n$ has stimulated the synthesis and study of the dielectric properties of 
various members of the RP series. Through the use of conventional ceramic processing 
techniques, single phase Sr$_{2}$TiO$_{4}$ ($n=1$),~\cite{Tilley,RP,McCarthy,Itoh,Wise} 
Sr$_{3}$Ti$_{2}$O$_{7}$ ($n=2$),~\cite{Tilley,McCarthy,Itoh,Wise,RP2} and 
Sr$_{4}$Ti$_{3}$O$_{10}$ ($n=3$),~\cite{McCarthy} were formed. Recently, thin films 
of the first five members of the RP series were grown by molecular beam epitaxy (MBE) 
on SrTiO$_{3}$ substrates.~\cite{Schlom1,Schlom2} Sr$_{2}$TiO$_{4}$, 
Sr$_{3}$Ti$_{2}$O$_{7}$, and Sr$_{4}$Ti$_{3}$O$_{10}$ were found to be nearly single phase
while Sr$_{5}$Ti$_{4}$O$_{13}$ and Sr$_{6}$Ti$_{5}$O$_{16}$ films showed noticeable 
antiphase boundaries and intergrowth defects.~\cite{Schlom2} The dielectric properties of 
the $n=1, 2$, and $\infty$ ceramic samples have been characterized by various 
groups.~\cite{Itoh,Wise,Paping,Itoh2} The dielectric properties of the $n=3$ and $4$ ceramic 
samples have also been studied but these were believed not to be single phase.~\cite{Wise}
It was found that SrTiO$_{3}$ had both the highest dielectric constant, $\epsilon_{r}$, 
and the largest TCF of the RP series. Conversely, Sr$_{2}$TiO$_{4}$ was found to have 
the lowest $\epsilon_{r}$ of the series but also the lowest TCF (an order of magnitude smaller 
than SrTiO$_{3}$) while having a dielectric loss comparable to  SrTiO$_{3}$.  
For thin films, Haeni {\it et al.}~\cite{Schlom2} performed low frequency and microwave 
experiments to measure the dielectric permittivity of Sr$_{2}$TiO$_{4}$ and of 
Sr$_{3}$Ti$_{2}$O$_{7}$, to be discussed below. 

Here we take the first step in understanding the dielectric behavior of the 
Sr$_{n+1}$Ti$_{n}$O$_{3n+1}$ Ruddlesden-Popper (RP) series from first principles by 
presenting our study of the structural and dielectric properties of Sr$_{2}$TiO$_{4}$ 
($n=1$) using density functional theory (DFT) structural optimization as well as 
density functional perturbation theory (DFPT).  DFT and DFPT have proved to be useful 
tools in investigating the structure, dynamical and dielectric properties of metals and 
insulators, including complex oxides (see Ref.~\onlinecite{Baroni} for a review).
In Section~\ref{sec:Method} we give the details of our DFT/DFPT calculations and 
describe various constraints that we imposed on the Sr$_{2}$TiO$_{4}$ structure. 
In Section~\ref{sec:results} the results of our calculations for the ground-state 
structural parameters, Born effective charges, zone-center phonons and electronic 
and static dielectric permittivity tensors at T=0 are presented and compared with experiment.  
In Section~\ref{sec:discuss} we discuss the sensitivity of the dielectric response to 
structural constraints, and through comparison with SrTiO$_{3}$, explore the key
role played by the Ti-O chain geometry~\cite{Ghosez98b} in the dielectric response
and its anisotropy in the RP phases. Finally, in Section~\ref{sec:Summary} we summarize 
our results and main conclusions.

\section{Method}
\label{sec:Method}
\subsection{First principles calculations}
\label{sec:firstprinciples}
First-principles density-functional calculations were performed within the local 
density approximation (LDA) as implemented in the ABINIT package.~\cite{abinit} The 
exchange-correlation energy is evaluated using the Teter rational polynomial fit to 
the Ceperley-Alder electron-gas data.~\cite{Ceperley80} Teter extended norm-conserving
pseudopotentials were used with Ti(3s 3p 4s 3d), Sr(4s 4p 5s), and O(2s 2p) levels 
treated as valence states. The electronic wavefunctions were expanded in plane waves 
up to a kinetic energy cutoff of 35 Ha.  Integrals over the Brillouin zone were approximated 
by sums on a $6 \times 6 \times 2$ mesh of special $k$-points.~\cite{Monkhorst76+}

\subsection{Structural Constraints}

We first performed full optimization of the lattice parameters and internal 
coordinates in the reported structure of Sr$_{2}$TiO$_{4}$,~\cite{RP} the 
body-centered tetragonal K$_{2}$NiF$_{4}$ structure (space group 
$I4/mmm (D^{17}_{4h}$), with the primitive unit cell containing one formula 
unit as shown in Fig.~\ref{fig:struct}. Ti atoms occupy Wyckoff position (2a), 
O$_{x}$ and O$_{y}$ atoms (4c), and Sr and O$_{z}$ atoms (4e), the latter with 
one free parameter (displacement along $\hat z$) each (see Table~\ref{table:wyckoff}). 
This yields the predicted structure under zero stress, roughly corresponding 
to bulk ceramic powders and relaxed epitaxial films. The structural relaxation 
was performed using a modified Broyden-Fletcher-Goldfarb-Shanno (BFGS) algorithm
~\cite{BFGSmod} to optimize the volume and atomic positions, followed by 
an optimization of the $c/a$ ratio. This procedure was repeated to ensure 
convergence. The residual Hellmann-Feynman forces were less than 2 meV/\AA. A 
second structure was considered to investigate the effects of epitaxial strain 
induced by the SrTiO$_{3}$ substrate on a fully coherent thin film. With the same 
space group, we fixed the in-plane lattice parameter to that of SrTiO$_{3}$ 
calculated within the present theory (a= 3.846\AA),~\cite{mySTO3} while optimizing 
the other structural parameters: the $c$ lattice constant and displacements along 
$[$001$]$. This was followed by ionic relaxation until the residual Hellmann-Feynman 
forces were less than 1 meV/\AA.

The next section will show that the fully optimized lattice parameters are 
underestimated relative to the experimental values, a common feature in LDA 
calculations.~\cite{VanOpinion} To investigate whether our calculations of the 
dielectric response of Sr$_{2}$TiO$_{4}$ are sensitive to this underestimation of 
the volume, two additional structures were considered. First, with the same space 
group, we optimized all internal coordinates fixing both the $a$ and $c$ lattice 
constants to values obtained by uniformly expanding the lattice constant of the 
in-plane constrained structure using the measured thermal expansion coefficient of 
SrTiO$_3$~\cite{Sai00,Lytle64} ($\alpha\approx$7x10$^{-6}$/K) and a temperature 
increase of 300 K. We also optimized internal coordinates fixing both the $a$ and 
$c$ lattice constants to the values experimentally obtained for the thin film sample. 
In both structures the ions were relaxed until the residual Hellmann-Feynman forces 
were less than 1 meV/\AA.

\subsection{Linear response calculations}
\label{sec:calculations}
Linear response methods provide an efficient means for computing quantities
that can be expressed as derivatives of the total energy $E$ with respect to
a perturbation, such as that produced by displacement $u_{i\alpha}$ of
an atom or a homogeneous electric field $\cal E$. Examples computed in this 
work include the force constant matrix elements
$${\partial^2E \over \partial u_{i\alpha}\partial u_{j\beta}}\vert_0$$ where 
$u_{i\alpha}$ is the displacement of atom $i$ in cartesian direction $\alpha$
from its position in the equililbrium crystal structure, the Born effective 
charge tensor $$Z^*_{i\alpha\beta}=-{\partial^2 E \over \partial d_{i\alpha} 
\partial {\cal E}_\beta}$$ where $d_{i\alpha}$ is the uniform displacement of 
the atomic sublattice $i$ in cartesian direction $\alpha$ from its position in 
the equilibrium unit cell, with derivatives taken in zero macroscopic field, 
and the electronic susceptibility tensor $$\chi_{\alpha\beta}=-{\partial^2 E 
\over \partial {\cal E}_\alpha \partial {\cal E}_\beta}$$ which is related to 
the electronic dielectric permittivity tensor $\epsilon_\infty$ by 
$\epsilon_\infty = 1 + 4 \pi \chi$. In this work, we compute these 
derivatives using the variational formulation of density-functional perturbation 
theory (DFPT), implemented in the ABINIT package.~\cite{abinit,gonze97,gonze97b}

The static dielectric permittivity tensor $\epsilon_0$ can be obtained directly 
from the quantities computed by DFPT. In general the static dielectric permittivity 
tensor of a nonpolar material can be written~\cite{Gonze01a,Gonze01b} as the sum 
of the electronic dielectric permittivity tensor $\epsilon_{\infty}$ and a sum of 
contributions $\Delta\epsilon_m$ from each of the zone-center polar modes $m$:
\begin{equation}
\epsilon^{0}_{\alpha \beta} = \epsilon^{\infty}_{\alpha\beta} +
\sum_{m}\Delta\epsilon_{m,\alpha\beta}
\label{eq:TotDIE}
\end{equation}
where $\Delta\epsilon_{m}$ is given by~\cite{Born,Mara}
\begin{equation}
\Delta\epsilon_{m,\alpha\beta}=
\frac{4 \pi e^{2}}{M_0\,V} 
\frac{\widetilde{Z}^{*}_{m\alpha}\,\widetilde{Z}^{*}_{m\beta}}{\omega_{m}^{2}}
\label{eq:ModDIE}
\end{equation}

Here, $V$ is the volume of the primitive unit cell and $M_{0}$ is a reference 
mass taken as 1 amu.  
$\frac{4 \pi e^{2}}{M_0\,V} \widetilde{Z}^{*}_{m\alpha}\widetilde{Z}^{*}_{m\beta}$ 
can be thought of as an effective plasma frequency, $\Omega_{p,m}^2$, of 
the $m^{th}$ normal mode,~\cite{ziman,d.y.smith} while $\omega_{m}$ is the 
frequency of vibration of normal mode $m$. $\widetilde{Z}^{*}_{m \alpha}$, 
which has been referred to as a mode effective charge,~\cite{Zhao,zhao.hf,
cockayne.burton} is given by
\begin{equation}
\widetilde{Z}^{*}_{m \alpha} = \sum_{i \gamma} \, Z^{*}_{\alpha\gamma}\left(i \right) \,
 \left(\frac{M_0}{M_i}\right)^{1/2} \, \xi_{m}\left(i \gamma\right)
\label{eq:Modchrg}
\end{equation}
where $\xi_{m}$ is the dynamical matrix eigenvector; the corresponding real space 
eigendisplacement of atom $i$ along $\beta$ is given by $U_{m}\left(i \beta\right) 
= \xi_{m}\left(i \beta\right)/M_i^{1/2}$.~\cite{modcharge.explain} Thus we see that 
large lattice contributions to the static dielectric permittivity tensor are expected 
if the relevant mode frequencies are very low and/or if the effective plasma frequencies 
are large (reflecting large mode effective charges).

This formalism has been applied in a first-principles context in numerous 
calculations of the zero-temperature static dielectric response (for a review, 
see Ref.~\onlinecite{Umeshreview}) in simple and complex oxides. For example, 
recent calculations for zircon (ZrSiO$_{4}$)~\cite{Gonze01a} and zirconia 
(ZrO$_{2}$)~\cite{Gonze01b,Zhao} yield good agreement with experiment.

\section{Results}
\label{sec:results}
\subsection{Crystal Structure of Sr$_{2}$TiO$_{4}$}
\label{sec:struct}
The reported crystal structure of Sr$_{2}$TiO$_{4}$ can also be viewed as a 
stacking of TiO$_{2}$ and SrO planes along [001] as shown in Fig.~\ref{fig:struct}(b), 
the stacking sequence being TiO$_{2}$-SrO-SrO where the 
second SrO layer is shifted with respect to the previous SrO layer by 
$\frac{1}{2}a_{0}$ along [110]. In previous work (Ref.~\onlinecite{noguera}), 
the internal structural parameters are presented as $u'$ and $v$, the distance 
between TiO$_{2}$ and SrO planes and between successive SrO planes, respectively, 
and $\delta$, the distance along $c$ between Sr and O in the same SrO layer, 
which quantifies the ``rumpling" of the layer.

In Table~\ref{table:parameters} we present the theoretical lattice parameters 
for the structures discussed in Section~\ref{sec:Method} and compare them
with the experimental values for ceramic samples~\cite{RP,Itoh,Burns87b} and for thin films 
epitaxially grown on SrTiO$_{3}$ (a=3.905\AA) substrates.~\cite{Schlom1,Schlom2}
The lattice constants of the fully relaxed structure, Table~\ref{table:parameters}(a), 
are smaller than the measured lattice constants of the ceramic powder. Specifically, 
the $a$ lattice parameter was calculated to be less than experiment by 1.5$\%$, 
which is within the error typically associated with the LDA, while the $c$ lattice 
parameter is underestimated by 2.2$\%$. The smaller value of $c$ measured for the 
thin film samples reduces this discrepancy to 1.1$\%$. There is considerable rumpling 
of the SrO layers, $\delta_{SrO}$ = 0.05$a_{0}$. This rumpling is such that the O$_{z}$ and Sr atoms 
move in a direction away from and towards the TiO$_{2}$ layers, resulting in a Ti-O$_{z}$ 
bond slightly larger (2.9$\%$) than the Ti-O$_{y}$ (or equivalently Ti-O$_{x}$) bond.
As pointed out by Noguera, Ruddlesden and Popper had assumed that this rumpling was 
equal to zero. The present study supports the previous calculation~\cite{noguera} by 
finding a nonzero rumpling. The experimental work of Venkateswaran {\it et al.}~\cite{Burns87b} 
also suggested a nonzero rumpling of the SrO layer ($\delta_{SrO}$=0.10),~\cite{burns.explain}
but found that the Ti-O$_{z}$ bond is shorter than the Ti-O$_{y}$ bond.

Next, we consider the structure in which we constrained the in-plane lattice constant to 
that of theoretical SrTiO$_{3}$ ($a$ = 3.846 \AA), Table~\ref{table:parameters}(b). 
This places the system under tensile in-plane stress equal to that of an experimental 
sample of Sr$_{2}$TiO$_{4}$ coherently matched to SrTiO$_{3}$ (lattice mismatch 0.6$\%$).
As can be seen in Table~\ref{table:parameters}, this has almost no effect on the 
optimized values of the other structural parameters, including $c$. 

Finally, we consider the optimized structural parameters for the two expanded structures.
The slight thermal expansion of the in-plane constrained structure has a correspondingly
slight effect on the internal parameters, while the expanded $c$ of the experimental 
thin film structure has a larger effect. We see that a roughly homogeneous expansion of the 
lattice, i.e. column (a)$\rightarrow$(d) and (b)$\rightarrow$(c), decreases the rumpling 
of the SrO layer.

\subsection{Born Effective Charge Tensors}
\label{sec:BECs}

In the Sr$_{2}$TiO$_{4}$ structure, the site symmetries of the Ti atom, occupying Wyckoff 
position 2a, and the O$_{z}$ and Sr atoms, both occupying Wyckoff position 4e, are 
tetragonal, while that of the O$_{x}$ and O$_{y}$ atoms, Wyckoff position 4c, is 
orthorhombic.  As a result, all $Z^{*}$'s are diagonal with two ($Z^{*}(Ti)$, 
$Z^{*}(O_{z})$, and $Z^{*}(Sr)$) and three ($Z^{*}(O_{y})$ and $Z^{*}(O_{x})$) independent 
components. Table~\ref{table:BECs} displays the diagonal components of the calculated Born 
effective charge tensors in Sr$_{2}$TiO$_{4}$ for the four structures considered. We see 
that the $Z^{*}$'s are relatively insensitive to the various volume and strain constraints 
imposed, consistent with what was shown for isotropic volume changes in BaTiO$_{3}$~\cite{Ghosez95} 
and for pure tetragonal strain in KNbO$_{3}$.~\cite{Wang96} Also, we note 
an anomalously large value for $Z^{*}_{xx}(Ti)=Z^{*}_{yy}(Ti)$ 
(nominal charge $+4$) and for $Z^{*}_{xx}(O_{x})=Z^{*}_{yy}(O_{y})$ (nominal charge $-2$), 
these being the components of the $Z^{*}$'s corresponding to motion parallel to the Ti-O bond 
along a direction in which the infinite Ti-O chains have been preserved. In contrast, the 
anomalous parts of $Z^{*}_{zz}(Ti)$ and $Z^{*}_{zz}(O_{z})$, which again correspond to motion 
parallel to the Ti-O bond, but along [001] where the Ti-O bonds do not form continuous chains, 
are found to be less than half of those along the continuous Ti-O chain.

\subsection{Phonon Frequencies at $\Gamma$}
\label{sec:phonons}
For the Sr$_{2}$TiO$_{4}$ structure, group-theoretical
analysis predicts that the 18 zone-center optic modes transform according to the following 
irreducible representations
\begin{equation}
\label{eqn:gamma_modes}
\Gamma_{optic} = 2 A_{1g} \oplus 2 E_{g}\oplus 3 A_{2u} \oplus 4 E_{u} \oplus  B_{2u}
\end{equation}
of which the A$_{1g}$ and E$_{g}$ modes are Raman active, the A$_{2u}$ and E$_{u}$
modes are infrared active, and the B$_{2u}$ mode is neither Raman nor infrared active.
By using projection operator methods, it can be shown that the A$_{1g}$, 
A$_{2u}$, and B$_{2u}$ modes involve motion along [001] while in the E$_{g}$ and 
E$_{u}$ modes, atoms move along [100] and [010]. A complete listing of the symmetry adapted
lattice functions is given in Refs.~\onlinecite{Burns88} and \onlinecite{Burns87}. 

Table~\ref{table:freqmod} displays our calculated frequencies and mode assignments for 
the four structures considered. Comparing the calculated phonon frequencies for the fully 
relaxed structure (a) with those measured by Burns {\it et al.} on ceramic samples,~\cite{Burns88} 
also given in Table~\ref{table:freqmod}, we find excellent agreement. The calculations 
allow us to confirm the mode assignments of the A$_{2u}(TO1)$ and E$_{u}(TO3)$: Burns' 
suggestion of assigning the lower mode at 242 cm$^{-1}$ A$_{2u}$ symmetry and the higher 
mode at 259 cm$^{-1}$ E$_{u}$ symmetry appears to be correct. In addition, Burns' 
assessment that the strong $LO$ feature at $\approx$440 cm$^{-1}$ in the measured 
reflectivity spectra was ``probably" due to both a A$_{2u}(LO)$ and a E$_{u}(LO)$ mode 
is consistent with our calculations, where we found A$_{2u}(LO)=$479 cm$^{-1}$ and 
E$_{u}(LO)=$451 cm$^{-1}$.~\cite{burns.explain.2} In fact an average of these two frequencies 
is within 2 cm$^{-1}$ of the quoted value 467 cm$^{-1}$. Finally, we obtain the frequencies 
of four modes that could not be separately identified in the experiment, namely the 
A$_{2u}(TO2)$ mode at 378 cm$^{-1}$, A$_{2u}(LO)$ mode at 252 cm$^{-1}$, the E$_{u}(TO4)$ 
mode at 611 cm$^{-1}$, and the B$_{2u}$ mode at 303 cm$^{-1}$. 

In Table~\ref{table:eigenmod} we give the normalized dynamical matrix eigenvectors
for the fully relaxed structure (a). We see that the lowest frequency A$_{2u}(TO1)$ 
and E$_{u}(TO1)$ modes involve Sr atoms moving against a fairly rigid TiO$_{6}$ 
octahedron, with larger deformation of the octahedron for the E$_{u}$ mode. The 
highest frequency A$_{2u}(TO3)$ and E$_{u}(TO4)$ modes each involve large distortions
of the oxygen octahedra with the apical oxygens moving against the planar oxygens in 
the A$_{2u}$ mode and a corresponding distortion for the E$_{u}$ mode. The calculated 
A$_{2u}(TO2)$ mode involves Ti atoms moving against a nearly rigid oxygen octahedron, 
giving rise to a relatively large mode effective charge. Finally, the E$_{u}(TO2)$ 
mode, characterized by Burns {\it et al.} as a weak infrared active mode, has 
displacements of O$_{y}$ and O$_{z}$ very close to those of the triply degenerate 
silent mode in the cubic perovskite. These patterns are substantially different from 
those obtained in model calculations of the lattice dynamics of K$_{2}$ZnF$_{4}$,~\cite{Rauh} 
used by Burns to characterize the eigenvectors of Sr$_{2}$TiO$_{4}$, especially for 
the highest frequency A$_{2u}(TO3)$ and E$_{u}(TO4)$ modes and the A$_{2u}(TO2)$ mode.

The effective plasma frequencies (Sec.~\ref{sec:calculations}) are reported 
in Table~\ref{table:ModChrg}. There is relatively little variation among the 
different modes of the same symmetry in the same structure, with $\Omega_{p,m}$ 
$\approx$400 cm$^{-1}$ and 800 cm$^{-1}$ for the A$_{2u}$ and E$_{u}$ respectively,
and even less variation with changes in the structure. The largest calculated 
effective plasma frequency is obtained for the A$_{2u}(TO2)$ mode. While the 
corresponding oscillator strength, which goes like $\Omega^2_{p,m}$, should make 
this the most prominent mode in a single-crystal infrared study, it was not 
separately identified in the ceramic sample of Burns. In ceramics, while the 
reststrahlen bands do not overlap for modes transforming as the same irreducible 
representation, the A$_{2u}$ modes can overlap the E$_{u}$ and vice versa. In fact 
we see that not only the A$_{2u}(TO2)$ mode, but also a second high-oscillator 
strength mode, E$_{u}(TO4)$, not identified in the measurements of Burns, lie in 
the middle of a wide reststrahlen band of the other symmetry type. As noted by 
Burns, this would undoubtedly complicate the interpretation of the reflectivity 
spectrum and may explain why neither of these modes were identified.

\subsection{Dielectric Permittivity Tensors}
\label{sec:DIE}
Here we present our calculation of the static dielectric permittivity tensor
$\epsilon_{0}$ for the structures considered in Table~\ref{table:parameters}. 
By the symmetry of the Sr$_{2}$TiO$_{4}$ structure we see that the tensors are 
diagonal and have two independent components, along directions parallel to and 
perpendicular to the TiO$_{2}$ layers. In Table~\ref{table:DIE} we display the 
results of our calculations for these three tensors ($\epsilon_{\infty}$, 
$\epsilon_{ionic}$, and $\epsilon_{0}$) for Sr$_{2}$TiO$_{4}$ under the various 
structural constraints previously discussed (see the caption of Table
~\ref{table:parameters}). It can be seen that the static dielectric permittivity 
tensor is quite anisotropic, with the in-plane components nearly 3 times as large 
as the component along [001]. 

To improve our understanding of this anisotropy we examine the contribution to 
$\epsilon_{ionic}$ from individual phonon modes. In Table~\ref{table:ModChrg} 
we show the effective plasma frequency, $\Omega_{p,m}$, and the contribution 
to the static dielectric permittivity for each IR-active phonon mode, equal
to $\frac{\Omega^{2}_{p,m} }{\omega_{m}^{2}}$. The A$_{2u}$ and the E$_{u}$ modes 
contribute to the components of $\epsilon_{0}$ along directions perpendicular and 
parallel to the TiO$_{2}$ layers, respectively. We see that the dominant mode 
contributing to the rather large anisotropy is the E$_{u}(TO1)$ mode, while the 
largest $\Omega_{p,m}$ is associated with the A$_{2u}(TO2)$ mode. In fact 
the effective plasma frequency of the A$_{2u}(TO2)$ mode is $\sim$40$\%$ larger than 
that of the E$_{u}(TO1)$ mode. The fact that the frequency of this A$_{2u}(TO2)$ 
mode is more than twice that of the soft E$_{u}(TO1)$ mode explains its relatively 
small contribution to the dielectric permittivity tensor.

Finally, to compare the calculated dielectric permittivity tensors with the
measured dielectric constant of the ceramic samples, $\epsilon_{r}$, we compute an
orientational average, $\epsilon_{average}$, of our calculated results. From 
Table~\ref{table:DIE} we see that $\epsilon_{average}= 38$ for the fully-relaxed structure
agrees quite well with $\epsilon_{r}= 34 - 38$ measured by the various groups.
The comparison with the thin film results is more problematic and will be discussed
in the next section.

\section{Discussion}
\label{sec:discuss}

In this section, we discuss the main features of the T=0 dielectric response of 
Sr$_2$TiO$_4$ and its relationship to that of SrTiO$_3$. We show that the 
anisotropy and sensitivity to changes in the lattice constants through epitaxial 
strain and thermal expansion can be understood by focusing on the Ti-O chains and 
chain fragments. Finally, we discuss the comparison of the results of our calculations 
to available experimental data.

As noted in the previous section, the static dielectric tensor of Sr$_2$TiO$_4$ 
is rather anisotropic, with the response in the planes of the layers about three 
times greater than the response perpendicular to the layers. This can be understood 
by analyzing the Ti-O chains. In SrTiO$_3$, the large dielectric response can 
be directly linked to the presence of infinite Ti-O chains running in all three 
cartesian directions, with anomalously large Born effective charges and low 
frequencies for modes with Ti displacing relative to O along the chain~\cite{Zhong94,lasota}
(see Table~\ref{table:srtio3}). In Sr$_{2}$TiO$_{4}$ the infinite Ti-O chains lying in the 
TiO$_{2}$ planes are preserved, while the Ti-O chains along [001] are broken into 
short O-Ti-O segments by the relative shift of the SrO-terminated perovskite slabs
(see Figure~\ref{fig:struct}(c)). 
Correspondingly, as can be seen in Table~\ref{table:BECs} the $Z^{*}$'s in 
Sr$_{2}$TiO$_{4}$ for Ti and O displacing along the infinite chains in the TiO$_2$ 
layers, $Z^{*}_{yy}(Ti)$ ($Z^{*}_{xx}(Ti)$) and $Z^{*}_{yy}(O_{y})$ ($Z^{*}_{xx}(O_{x})$), 
are only slightly smaller than those of SrTiO$_3$.~\cite{bec.explain} In contrast, 
the breaking of the Ti-O chains into O-Ti-O segments along [001] has a dramatic 
effect on the anomalous component of the Ti and O Born effective charges for motion 
along the segment, $Z^{*}_{zz}(Ti)$ and $Z^{*}_{zz}(O_{z})$, reducing them by over 
a factor of 2. While it is true that the local anisotropy caused by the somewhat 
larger Ti-O$_{z}$ bond length compared with the Ti-O$_{y}$ bond length will also 
reduce the corresponding Born effective charges, this is a much weaker effect as 
suggested by the following calculation. We fixed the Ti-O$_{z}$ distance to that of 
the Ti-O$_{y}$ bond length of structure (b) (where the in-plane plane lattice 
parameter was fixed to the lattice constant of SrTiO$_{3}$), creating a structure 
whereby the Ti-O distances are equal in all three cartesian directions. The Born 
effective charges along [001] were found to increase, but only slightly 
($Z^{*}_{zz}(Ti)$=5.29 and $Z^{*}_{zz}(O_{z})$=-3.80).

This reduction of the anomalous component of the Born effective charges is quite
similar to the reduction of the $Z^{*}$'s observed in BaTiO$_{3}$ due to the 
ferroelectric transition. Indeed, Ghosez {\it et al.}~\cite{Ghosez95,Ghosez98b} 
found that $Z^{*}(Ti)=7.29$ and $Z^{*}(O_{\|}) =-5.75$ in cubic BaTiO$_{3}$ while 
in the tetragonal phase they found $Z^{*}_{yy}(Ti)=6.94$, $Z^{*}_{yy}(O_{y})=-5.53$, 
$Z^{*}_{zz}(Ti)=5.81$ and $Z^{*}_{zz}(O_{z})=-4.73$.  It was explained that this
reduction of the $Z^{*}$'s going from the cubic to the tetragonal phase resulted from 
the displacement of the Ti atom along the ferrroelectric axis, resulting in a series 
of long-short Ti-O bonds thereby ``breaking" the Ti-O chains.~\cite{Ghosez98b} In 
Sr$_{2}$TiO$_{4}$ this breaking of the Ti-O chains can be thought of as being caused not 
by alternating Ti-O bond lengths along a given direction, but by the shift of the 
SrO-terminated SrTiO$_{3}$ slabs, resulting in a SrO-SrO anti-phase boundary perpendicular 
to [001] as discussed in Sec.~\ref{sec:struct}. Finally, this breaking of the chains 
in Sr$_{2}$TiO$_{4}$ is arguably a much ``stronger" effect than the cubic to tetragonal 
transition in BaTiO$_{3}$, as evidenced by the greater reduction of the Born effective 
charges along [001].

The breaking of the [001] Ti-O chains into segments lowers the dielectric response
relative to SrTiO$_3$ also by stiffening the low-frequency mode associated with relative
displacement of Ti and O along the chains. The in-plane $E_u$ (TO1) mode in Sr$_2$TiO$_4$ 
for the fully relaxed structure (a) is at 148 cm$^{-1}$ (134 cm$^{-1}$ for the structure
with in-plane lattice constant fixed to SrTiO$_{3}$ (b)) comparable to our calculated 
value of 103 cm$^{-1}$ for SrTiO$_3$,  while the corresponding
out-of-plane $A_{2u}$ (TO2) mode is at 378 cm$^{-1}$ for structure (a) (391 cm$^{-1}$ for
structure (b)). As we discussed in the previous section, the large anisotropy is
almost completely accounted for by the large $\Delta\epsilon_m$ of the in-plane $E_u$ (TO1) mode
as shown in Table ~\ref{table:ModChrg}.

The topology of the Ti-O chains thus appears to be much more important than 
the bond lengths and the details of the geometry. Indeed, our calculations 
show that the dielectric permittivity tensor of Sr$_{2}$TiO$_{4}$ is not 
very sensitive to an in-plane tensile strain. From Table~\ref{table:DIE} we 
see that $\epsilon_{average}$ for the fully-relaxed structure (a) and for 
the structure with the in-plane lattice constant constrained to the theoretical 
SrTiO$_{3}$ value (b) only differ by $\sim$15$\%$. Also, we see that this 
difference originates from the component of $\epsilon_{ionic}$ along in-plane 
directions while the lattice contribution to the dielectric permittivity tensor 
along [001] remains relatively constant. In particular, we see that $\epsilon_{ionic}$ 
along directions in which the Ti-O chains were preserved is relatively more 
sensitive to a fixed strain than along directions in which these chains were 
broken. In fact this sensitivity can be attributed to a single E$_{u}(TO1)$ 
mode where it can be seen that the sum of the contributions to $\epsilon_{ionic}$ 
from the remaining E$_{u}(TO)$ modes remains nearly constant for all the structures 
considered. As was pointed out in Sec.~\ref{sec:DIE} this E$_{u}(TO1)$ mode also 
contributes predominately to the rather large anisotropy of the static dielectric 
permittivity tensor. Thus, the difference in response between fully coherent and 
fully relaxed films is expected to be relatively insignificant, unlike the case 
of epitaxial thin films of SrTiO$_{3}$, in which the induced strain would influence 
the lattice dynamics and presumably could have significant effect on the dielectric 
properties.~\cite{Hyun,Xi.james} In addition, this insensitivity should reduce the TCF, 
especially for $\epsilon_{33}$.

Anisotropy of the dielectric tensor has not been previously experimentally determined
in Sr$_2$TiO$_4$, though measurements have been carried out for single crystals of some 
isostructural oxides CaYAlO$_{4}$, CaNdAlO$_{4}$, and LaSrAlO$_{4}$.~\cite{shannon98} 
In the latter case, the anisotropy is very small (and in the case of LaSrAlO$_{4}$, 
the out-of-plane component of the dielectric tensor, $\epsilon_{33}$= 20.02, is slightly 
greater than the in-plane component $\epsilon_{11}$= $\epsilon_{22}$=16.81). The absence 
of a large in-plane response in these materials highlights the unique physics of the Ti-O chain.

The anisotropy of the dielectric tensor is an important factor to be taken into 
consideration when comparing the computed dielectric response with experiment, 
and indeed in correctly interpreting experimental measurements. For ceramics, 
anisotropy of the single crystal response can give a range of values for the 
measured dielectric response, depending on the size, shape and interaction of 
the grains. In the single crystal films, the anisotropy complicates the 
interpretation of the microwave microscope method of determining the dielectric 
response. The effective dielectric constant obtained by this technique is in fact 
an average of the dielectric tensor components determined by the field distribution 
near the probe, which is itself determined by the anisotropic dielectric response. 
While an exact value requires detailed modeling of the field configuration, the 
expectation is that the measured value should be an average weighted in favor of 
the high dielectric response components. Thus, the value $\epsilon_{eff}$ = 44 
obtained by Haeni~\cite{Schlom2} is not $\epsilon_{33}$, but is a weighted average 
of the three components, in accordance with our calculations. The agreement of this 
$\epsilon_{eff}$ with the low-frequency electronic measurement is apparently 
coincidental. The latter measurement determines $\epsilon_{33}$, but at low frequencies 
it is not unusual for additional extrinsic contributions to raise the value of the 
dielectric response.

\section{Summary}

Our calculation of the static dielectric tensor of Sr$_2$TiO$_4$ provides valuable
information, not previously available, about the anisotropy of the response.
This anisotropy is key to interpreting the available experimental data on the
dielectric response of ceramic and thin film samples. We identify a dominant role 
played by Ti-O chains in the lattice contribution to the response, which allows us to 
understand the anisotropy. This picture should generalize to higher RP phases of Sr-Ti-O, 
and further could be useful in tailoring the dielectric response of other perovskite-based 
systems.

\label{sec:Summary}

\acknowledgments
The authors would like to acknowledge valuable discussions with Darrel Schlom, 
David Vanderbilt, and Morrel Cohen. This work was supported by NSF-NIRT DMR-0103354.

\begin{table*}
\caption{Wyckoff positions for Sr$_{2}$TiO$_{4}$.}
\begin{ruledtabular}
\begin{tabular}{lll}
Atom&Wyckoff position and point symmetry&Coordinates\\
\hline
Ti& (2a) $4/mmm$ &0,0,0\\
O$_{x}$, O$_{y}$ & (4c) $mmm$ &$1\over2$,0,0; 0,$1\over2$,0\\
O$_{z}$ & (4e) $4mm$ &0,0,$\pm$z$_{Oz}$\\
Sr & (4e) $4mm$ &0,0,$\pm$z$_{Sr}$\\
\end{tabular}
\end{ruledtabular}
\label{table:wyckoff}
\end{table*}

\begin{table*}
\caption{Structural parameters of Sr$_{2}$TiO$_{4}$.
(a) Fully relaxed cell. (b) Relaxation along c-axis, with $a$ lattice constant fixed at
theoretical SrTiO$_{3}$. (c) Relaxation with lattice constants fixed at values corrected 
for thermal expansion. (d) Lattice constants fixed at experimental (Ref.~\protect\onlinecite{Schlom1}).
Lattice constants, bond lengths, and internal parameters are in \AA.}
\begin{ruledtabular}
\begin{tabular}{lcccccccccccc}
&\multicolumn{3} {c}{Experimental}  &&\multicolumn{4} {c}{Present Theory}
& &\multicolumn{3} {c}{Previous Theory}\\ \cline{2-4} \cline{6-9} \cline{11-13}
&\mbox{Ceramic}\footnotemark[1]& \mbox{Ceramic}\footnotemark[2]
&\mbox{Thin Film}\footnotemark[3] & & \mbox{(a)} & \mbox{(b)}
& \mbox{(c)} & \mbox{(d)} & & \mbox{Ref.~\protect\onlinecite{Suzuki}} \footnotemark[4]
&\mbox{Ref.~\protect\onlinecite{noguera}} \footnotemark[5]
&\mbox{Ref.~\protect\onlinecite{McCoy97}} \footnotemark[6]\\
\hline
Lattice Constants \\
\mbox{$a$}     &3.88  & 3.88   &3.88   & &3.822   &3.846   &3.855   &3.88  & &3.942 &4.0 &3.892\\
\mbox{$c$}    &12.60  &       &12.46   & &12.32   &12.27   &12.30   &12.46 & &12.56 &12.68 &12.70\\
\mbox{$c/a$}  &3.247  &        &3.211           & &3.223   &3.190   &3.190   &3.211 & &3.186 &3.17 &3.263\\
Bond Lengths \\
\mbox{Ti-O$_{x}$} & &1.94 & & &1.911   &1.923   &1.928   &1.940 & &&2.0\\
\mbox{Ti-O$_{z}$} & &1.92 & & &1.967   &1.960   &1.962   &1.974 & &&2.0\\
\mbox{Sr-O$_{z}$} & &2.56 & & &2.402   &2.397   &2.406   &2.455 & &&2.47\\
Internal Parameters \\
\mbox{$\delta_{\rm SrO}$}& &     & & &0.176 &0.182 &0.180 &0.173 & &&0.12\\
\mbox{u'}                & &     & & &1.879 &1.869 &1.872 &1.888 & &&1.94\\
\mbox{v}                 & &     & & &2.402 &2.397 &2.406 &2.455 & &&2.47\\
Reduced Coordinates\\
\mbox{z$_{Oz}$}          & &     & & &0.160 &0.160 &0.160 &0.158 & &&0.158\\
\mbox{z$_{Sr}$}          & &     & & &0.355 &0.355 &0.355 &0.355 & && 0.353\\
\end{tabular}
\end{ruledtabular}
\label{table:parameters}
\footnotetext[1]{Ref.~\onlinecite{Tilley,RP,McCarthy,Itoh}}
\footnotetext[2]{Ref.~\onlinecite{Burns87b}}
\footnotetext[3]{Ref.~\onlinecite{Schlom1}}
\footnotetext[4]{ab initio, DFT, GGA}
\footnotetext[5]{semiempirical Hartree-Fock}
\footnotetext[6]{empirical atomistic simulations}
\end{table*}

\begin{table*}
\caption{Nonzero components of the calculated Born effective charge tensors Sr$_{2}$TiO$_{4}$.}
\begin{ruledtabular}
\begin{tabular}{lcccccccccccccccccc}
&\multicolumn{3} {c}{(a)}  &&&\multicolumn{3} {c}{(b)}&&&\multicolumn{3} {c}{(c)}&&&
\multicolumn{3} {c}{(d)}\\
Atom&Z$^{*}_{xx}$&Z$^{*}_{yy}$&Z$^{*}_{zz}$&&&Z$^{*}_{xx}$&Z$^{*}_{yy}$&Z$^{*}_{zz}$&&&
Z$^{*}_{xx}$&Z$^{*}_{yy}$&Z$^{*}_{zz}$&&&Z$^{*}_{xx}$&Z$^{*}_{yy}$&Z$^{*}_{zz}$\\
\hline
Ti        &6.96&  6.96&  5.14&&& 6.89&  6.89&  5.20&&& 6.88&  6.88&  5.20&&&
                      6.88&  6.88&  5.15 \\
O$_{y}$    &-2.04& -5.46& -1.56&&& -2.05& -5.40& -1.55&&&  -2.04& -5.40& -1.55&&&
                      -2.04& -5.42& -1.52 \\
Sr       &2.37&  2.37&  2.75&&&   2.36&  2.36&  2.76&&&  2.36&  2.36&  2.75&&&
                       2.37&  2.37&  2.72 \\
O$_{z}$   &-2.09& -2.09& -3.74&&&   -2.08& -2.08& -3.79&&&  -2.07& -2.07& -3.78&&&
                      -2.08& -2.08& -3.75 \\
\end{tabular}
\end{ruledtabular}
\label{table:BECs}
\end{table*}

\begin{table}
\caption{Phonon frequencies (cm$^{-1}$) and mode assignments for Sr$_{2}$TiO$_{4}$.
Symmetry labels follow the convention of Ref.~\onlinecite{burns.glazer}.
Experimental values are from Ref.~\protect\onlinecite{Burns88}.}
\begin{ruledtabular}
\begin{tabular}{lccccc}
\mbox{Mode} &\mbox{Expt.} &\mbox{(a)}& \mbox{(b)}& \mbox{(c)} & \mbox{(d)}\\
\hline
Raman & & & & & \\
$A_{1g}$  &205 &216 &217 &214 &200\\
$A_{1g}$  &578 &588 &594 &588 &562\\
$E_{g}$   &124 &121 &118 &115 &106\\
$E_{g}$   &286 &271 &268 &266 &263\\
Infrared & & & & & \\
\mbox{$A_{2u}$ (TO1)}  &242  &231  &231  &227  &206\\
\mbox{$A_{2u}$ (TO2)}  &     &378  &391  &389  &368\\
\mbox{$A_{2u}$ (TO3)}  &545  &499  &501  &496  &478\\
\mbox{$A_{2u}$ (LO)}  &     &252  &253  &249  &233\\
\mbox{$A_{2u}$ (LO)}  &467  &479  &482  &480  &472\\
\mbox{$A_{2u}$ (LO)}  &683  &684  &692  &687  &658\\
\mbox{$E_{u}$ (TO1)}   &151  &148  &134  &129  &117\\
\mbox{$E_{u}$ (TO2)}   &197  &218  &211  &208  &198\\
\mbox{$E_{u}$ (TO3)}   &259  &246  &247  &244  &230\\
\mbox{$E_{u}$ (TO4)}   &     &611  &590  &581  &554\\
\mbox{$E_{u}$ (LO)}  &182  &184  &180   &177  &168\\
\mbox{$E_{u}$ (LO)}  &239  &227  &225   &223  &216\\
\mbox{$E_{u}$ (LO)}  &467  &451  &450   &448  &443\\
\mbox{$E_{u}$ (LO)}  &727  &789  &766   &756  &727\\
Silent & & & & & \\
$B_{2u}$  & &303 &310 &310 &303\\
\end{tabular}
\end{ruledtabular}
\label{table:freqmod}
\end{table}

\begin{table*}
\caption{Dynamical matrix eigenvectors, $\xi_{m}$, of Sr$_{2}$TiO$_{4}$
for the fully relaxed structure with a=3.822\AA\ and c=12.32\AA.
The corresponding eigendisplacement in real space can be obtained by
dividing each value by the appropriate mass factor $\sqrt{M_{i}}$.
Modes $A_{1g}$, $A_{2u}$, and $B_{2u}$ involve motion along [001], while
the two-fold degenerate modes $E_{g}$ and $E_{u}$ involve motion along [100] and
equivalently along [010].}
\begin{ruledtabular}
\begin{tabular}{lcccccccc}
\ & & \multicolumn{7}{c}{Eigenmodes} \\ \cline{3-9}
\mbox{Mode} &\mbox{Frequency}& \mbox{Ti}& \mbox{O$_{y}$} & \mbox{O$_{x}$}
 & \mbox{Sr(1)}& \mbox{Sr(2)}& \mbox{O$_{z}$(1)}& \mbox{O$_{z}$(2)} \\
\hline
Raman & & & & & & & &\\
$A_{1g}$  &216 &0   &0   &0   &-0.71   & 0.71   & 0.03   &-0.03\\
$A_{1g}$  &588 &0   &0   &0   &-0.03   & 0.03   &-0.71   & 0.71\\
$E_{g}$   &121 &0   &0   &0   &-0.70   & 0.70   &-0.06   & 0.06\\
$E_{g}$   &271 &0   &0   &0   & 0.06   &-0.06   &-0.70   & 0.70\\
Infrared & & & & & &\\
\mbox{$A_{2u}$ (TO1)}  &231   & 0.48   & 0.38   & 0.38   &-0.44   &-0.44   & 0.23   & 0.23\\
\mbox{$A_{2u}$ (TO2)}  &378   &-0.77   & 0.29   & 0.29   & 0.01   & 0.01   & 0.34   & 0.34\\
\mbox{$A_{2u}$ (TO3)}  &499   & 0.11   &-0.46   &-0.46   &-0.07   &-0.07   & 0.53   & 0.53\\
\mbox{$E_{u}$ (TO1)}  &148    & 0.14   & 0.11   & 0.24   &-0.37   &-0.37   & 0.56   & 0.56\\
\mbox{$E_{u}$ (TO2)}  &218    &-0.77   &-0.29   &-0.14   & 0.24   & 0.24   & 0.31   & 0.31\\
\mbox{$E_{u}$ (TO3)}  &246    & 0.45   &-0.86   &-0.19   & 0.02   & 0.02   & 0.09   & 0.09\\
\mbox{$E_{u}$ (TO4)}  &611    & 0.16   & 0.32   &-0.91   & 0      & 0      & 0.14   & 0.14\\
Silent & & & & & & & &\\
$B_{2u}$  &303 &0   & 0.71   &-0.71     &    0   &      0       &  0        & 0\\
\end{tabular}
\end{ruledtabular}
\label{table:eigenmod}
\end{table*}

\begin{table}
\caption{Effective plasma frequency, $\Omega_{p,m}$ (cm$^{-1}$), 
and mode contribution to the static dielectric tensor, 
$\Delta\epsilon_{m}$, in Sr$_{2}$TiO$_{4}$.}
\begin{ruledtabular}
\begin{tabular}{lccccccccccc}
 &\multicolumn{2}{c}{(a)}& &\multicolumn{2}{c}{(b)}&
&\multicolumn{2}{c}{(c)}& &\multicolumn{2}{c}{(d)}\\
\mbox{Mode}& $\Omega_{p,m}$ & \mbox{$\Delta\epsilon_{m}$}&
& $\Omega_{p,m}$ & \mbox{$\Delta\epsilon_{m}$}&
& $\Omega_{p,m}$ & \mbox{$\Delta\epsilon_{m}$}&
& $\Omega_{p,m}$ & \mbox{$\Delta\epsilon_{m}$}\\
\hline
\mbox{$A_{2u}$(TO1)} &457 &3.93 & &446  &3.71 & &449 &3.93 & &474 &5.28\\
\mbox{$A_{2u}$(TO2)} &1061 &7.86& &1066 &7.42 & &1073 &7.62& &1083 &8.66\\
\mbox{$A_{2u}$(TO3)} &427 &0.73 & &447  &0.80 & &409 &0.68 & &232 &0.24\\
\mbox{$E_{u}$(TO1)}  &752 &25.9 & &809  &36.44& &829 &41.00& &888 &57.24\\
\mbox{$E_{u}$(TO2)}  &473 &4.69 & &491  &5.41 & &507 &5.95 & &586 &8.80\\
\mbox{$E_{u}$(TO3)}  &794 &10.4 & &745  &9.08 & &723 &8.78 & &617 &7.21\\
\mbox{$E_{u}$(TO4)}  &812 &1.77 & &759  &1.65 & &737 &1.61 & &674 &1.48\\
\end{tabular}
\end{ruledtabular}
\label{table:ModChrg}
\end{table}

\begin{table*}
\caption{Nonzero components of the electronic and the ionic dielectric tensors for 
Sr$_{2}$TiO$_{4}$, ($\epsilon_{xx}, \epsilon_{yy}, \epsilon_{zz}$). Experimental measurements on
thin films (Ref.~\protect\onlinecite{Schlom2}) and on ceramic samples
(Ref.~\protect\onlinecite{Itoh,Wise,Paping}) were conducted at room temperature except
for $\dagger$ (T=15K.)}
\begin{ruledtabular}
\begin{tabular}{lcccccccc}
&  \multicolumn{4}{c}{Experimental} & & & &\\ \cline{2-5}
&Ref.~\protect\onlinecite{Schlom2}
&Ref.~\protect\onlinecite{Paping}
&Ref.~\protect\onlinecite{Itoh}
&Ref.~\protect\onlinecite{Wise}
& \mbox{(a)} &\mbox{(b)}& \mbox{(c)}& \mbox{(d)}\\
\hline
$\epsilon^{\infty}$ & & & & &( 5.09  5.09  4.81) &( 5.08  5.08  4.82) &( 5.07  5.07  4.82)
                    &( 5.08  5.08  4.79)\\
$\epsilon_{ionic}$  & & & & &(42.8  42.8  12.5)  &(52.6  52.6  11.9)  &(57.3  57.3  12.2)
                    &(74.7  74.7  14.2)\\
$\epsilon_{0}$      & & & & &(48    48    17)    &(58    58    17)  &(62  62  17)
                    &(80  80  19)\\
$\epsilon_{average}$ &$44\pm4$ &38 &34, 37$^\dagger$&37 &38   &44   &47   &60\\
\end{tabular}
\end{ruledtabular}
\label{table:DIE}
\end{table*}

\begin{table}
\caption{The structural and dielectric properties of SrTiO$_{3}$ calculated
within present theory. (6x6x6 grid, 45 Ha.) }
\begin{ruledtabular}
\begin{tabular}{ll}
Lattice Constant&  Phonons (cm$^{-1}$)\,\,\,\,\,$\omega_m$\,\,\,\,\,$\Omega_{p,m}$\\
a= 3.846 \AA&  A$_{2u}$(TO1)\,\,\,\,\,\,\,\,\,\,\,\,\,\,\,\,\,\,\,\,103\,\,\,\,\,1275\\
Born Effective Charge& A$_{2u}$(TO2)\,\,\,\,\,\,\,\,\,\,\,\,\,\,\,\,\,\,\,\,189\,\,\,\,\,\,\,\,827\\
$Z^{*}(Ti)$\,=\,\,\,\,\,$7.26$& A$_{2u}$(TO3)\,\,\,\,\,\,\,\,\,\,\,\,\,\,\,\,\,\,\,\,587\,\,\,\,\,\,\,\,946\\
$Z^{*}(Sr)$\,=\,\,\,\,\,$2.55$& A$_{2u}$(LO1)\,\,\,\,\,\,\,\,\,\,\,\,\,\,\,\,\,\,\,\,166\\
$Z^{*}(O_{\bot})$=$-2.04$ & A$_{2u}$(LO2)\,\,\,\,\,\,\,\,\,\,\,\,\,\,\,\,\,\,\,\,451\\
$Z^{*}(O_{\|})$\,=$-5.72$& A$_{2u}$(LO3)\,\,\,\,\,\,\,\,\,\,\,\,\,\,\,\,\,\,\,\,822\\
Dielectric Constant& \\
$\epsilon^{\infty}= 6.2$,\,\,\,\,\,$\epsilon_{0}=181$&\\
Phonon Eigenvector&$\{\,\,\,u_{Ti}, \,\,\,\,\,u_{Sr}, \,\,u_{O\parallel}, u_{O\perp}, u_{O\perp}\}$\\
$A_{2u}(TO1)$&\{-0.51, -0.14, 0.37, 0.54, 0.54\} \\
\end{tabular}
\end{ruledtabular}
\label{table:srtio3}
\end{table}


\begin{thebibliography}{10}
\bibitem{mySTO3}
Calculations were performed on $6\times6\times6$ grid at an energy cutoff of 45 Ha,
following the approach described in Section~\ref{sec:firstprinciples}.
\bibitem{modcharge.explain}
The mode effective charge, if defined as the constant of proportionality
between the induced polariaztion and a real-space eigendisplacement is given by
$\sum_{i \gamma}$ Z$^{*}_{\alpha\gamma}\left(i \right)U_{m}\left(i \gamma\right)$
/[$\sum_{i \gamma}$ $U^*_{m}\left(i \gamma\right)$
$U^*_{m}\left(i \gamma\right)$]$^{1\over2}$.
\bibitem{burns.explain} This assumes a c-axis lattice parameter (c=12.60\AA) equal 
to that previously reported in the literature for ceramic samples. Note, the c-axis lattice 
parameter is missing from their paper.
\bibitem{burns.explain.2}
In Table II of the Ref.~\cite{Burns88}, the A$_{2u}$ mode and the E$_{u}$ mode 
given rise to the strong LO feature at $\approx$ 440 cm$^{-1}$ was each assigned 
a value of 467 cm$^{-1}$.
\bibitem{bec.explain}
The hybridization of the Ti 3$d$ and O 2$p$ states has been shown to be the origin
of the anomalous Born effective charges (BECs) in the perovskites~\cite{Posternak94}.
Because one expects this hybridization, or the rate of change of this hybridization 
~\cite{Ghosez95}, to depend on the length of the Ti-O bond we should compare BECs 
in SrTiO$_{3}$ and Sr$_{2}$TiO$_{4}$ at the same bond length.
\end{thebibliography}
\end{document}